\documentclass[aps,twocolumn,prb,preprintnumbers,amsmath,amssymb,superscriptaddress,floats]{revtex4}

\usepackage{graphicx}
\usepackage{dcolumn}
\usepackage{bm}
\usepackage{hyperref}

\begin{document}

\title{Dynamical study of phase fluctuations and their critical slowing down in amorphous superconducting films}

\author{Wei Liu}
\affiliation{Department of Physics and Astronomy, Johns
Hopkins University, Baltimore, MD 21218 }

\author{Minsoo Kim}
\affiliation{Department of Physics, University at Buffalo-SUNY, 239 Fronczak Hall, Buffalo, NY 14260}

\author{G. Sambandamurthy}
\affiliation{Department of Physics, University at Buffalo-SUNY, 239 Fronczak Hall, Buffalo, NY 14260}

\author{N.P. Armitage }

\affiliation{Department of Physics and Astronomy, Johns
Hopkins University, Baltimore, MD 21218 }

\date{\today}

\begin{abstract}

We report a comprehensive study of the complex AC conductance of amorphous superconducting InO$_x$ thin films.  We measure the explicit frequency dependency of the complex conductance and the phase stiffness over a range from 0.21 to 15 GHz at temperatures down to 350 mK using a novel broadband microwave Corbino spectrometer.   The dynamic ac measurements are sensitive to the temporal correlations of the superconducting order parameter in the fluctuation range above $T_c$.  Among other aspects, we explicitly demonstrate the critical slowing down of the characteristic fluctuation rate on the approach to the superconducting state and show that its behavior is consistent with vortex-like phase fluctuations and a phase-ordering scenario of the transition.

\end{abstract}



\maketitle
\section{Introduction}
The remarkable properties of superconductors and superfluids arise from the macroscopic quantum-mechanical coherence of their complex order parameter (OP), $\Psi = \Delta e^{i\phi}$.  In conventional superconductors, fluctuations of the OP amplitude and phase occur in temperature regions only infinitesimally close to $T_c$.  In contrast, in disordered materials with reduced dimensionality, the situation may be considerably different.  Their low superfluid density gives a small phase stiffness and  phase fluctuations that may be particularly soft.\cite{Emery95a}  In such systems, phase plays the role of a dynamic variable and may result in a situation where the transition results from a phase disordering of the order parameter while its amplitude remains finite.  Effects such as zero resistivity are lost when the phase is no longer ordered on all lengths.   However, phase correlations may remain over finite length and time scales resulting in significant precursor effects above $T_c$.

In strictly two dimensions (2D), such a transition has been proposed \cite{Beasley79a,Halperin79a} to be of the Kosterlitz-Thouless-Berezinskii (KTB) variety, as in $^4$He films.\cite{Berezinskii72a,Kosterlitz73a,Bishop78a,Minnhagen87a}  In such a transition, thermally excited free vortices are not possible below the transition temperature $T_{\text{KTB}}$ as the vortex-antivortex binding energy increases logarithmically with separation.  However, above $T_{\text{KTB}}$ it becomes entropically favorable for vortices to unbind.  Vortex pairs with the largest separation unbind first and the phase stiffness  measured in the long-length and low-frequency limit suffers a discontinuous drop.  Vortex unbinding reduces the global phase stiffness and renders the system increasingly susceptible to further vortex proliferation.  At temperatures just above $T_{\text{KTB}} $ such systems can be described as a two-component vortex plasma and may be realizations of the 2D XY model.

Because free vortices are the topological defects of the phase field, their spacing plays the role of a Ginzburg-Landau correlation length $\xi$, which diverges as $T \rightarrow T_{\text{KTB}}$.  The role of free vortices as topological defects and their finite energy cost give an exponentially activated vortex density ($n_F \propto 1/\xi^2$).  Asymptotically close to the transition, this results in an unusually stretched exponential dependence of $\xi$ on temperature, $\xi \sim e^{\sqrt{T'/(T - T_{\text{KTB}}) }}$, which is in stark contrast to the power laws typically expected near continuous phase transitions.\cite{Kosterlitz73a}  Similar dependence is expected in the ``critical slowing down" of the phase correlation time $1/\Omega$, which in a vortex plasma, is proportional to the time $\xi^2/D$ required to diffuse the intervortex spacing (where $D$ is the vortex diffusion constant).\cite{Halperin79a}

 \begin{figure*}[t]
\begin{center}
\includegraphics[width=2\columnwidth,angle=0]{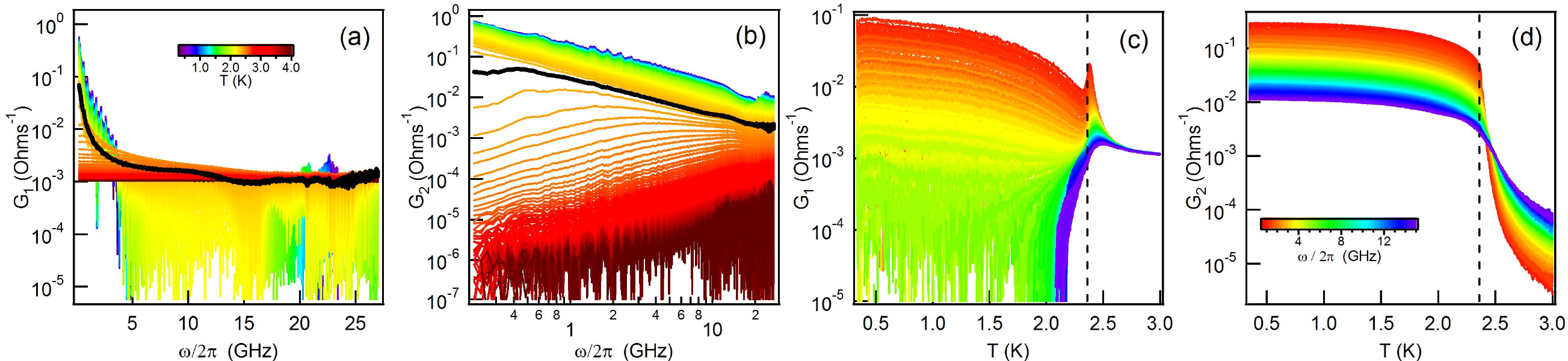}
\caption{(Color online) (a) and (b): frequency dependence of the real and imaginary conductances in the ranges $\omega/2 \pi = 0.21-27$ GHz and $T = 0.35-4$ K. A color scale representing different temperatures is displayed in (a). The black curves are the conductances at $T_c$. The features in (a) at about 22 GHz are residual features imperfectly removed during calibration. (c) and (d):  temperature dependence of the real and imaginary conductances in the frequency range $\omega/2 \pi = 0.660-15$ GHz.  A color scale representing different frequencies is displayed in (d). The dashed black lines mark $T_c = 2.36 $K.}
\label{conductivity}
\end{center}
\end{figure*}

Although the conventional wisdom is that a KTB-like transition occurs in ultrathin superconducting films,\cite{Beasley79a} the issue is still in fact controversial.  For instance, it has been proposed that, unlike $^4$He films, superconducting films are in a regime of low core energy (the high ``fugacity" limit), that causes the transition to acquire a nonuniversal character or even be first order.\cite{Minnhagen87a,Lee90a,Yu96a}   There has been a great deal of work looking for KTB physics in the linear and nonlinear dc transport characteristics of thin films superconductors.\cite{FioryPRB83,Epstein82a,KapitulnikPRB92}  But it is not clear to what extent these experiments are influenced by inhomogeneous broadening \cite{Benfatto09a} and if even KTB physics would be detectable in such experiments.\cite{Minnhagen87a}  In contrast, finite-frequency measurements can directly probe temporal correlations and can be explicitly sensitive to phase fluctuations right above $T_c$.  Important information has been gained from measurements at discrete frequencies,\cite{FioryPRB83,CranePRB07a,CranePRB07b,kitano07a} but only true spectroscopic measurements can give important information concerning critical slowing down.

In this paper, we present a comprehensive study of the complex ac conductance of effectively 2D amorphous superconducting InO$_x$ films.  We make use of our recent development of a broadband Corbino microwave spectrometer, which can measure the explicit frequency dependence of the complex conductance of thin films over a range from $0.21- 15$ GHz at temperatures down to 350 mK.  These unique measurements allow true spectroscopy in the microwave range at low temperatures.   We explicitly measure the temporal correlations of the fluctuation superconductivity and demonstrate the manner in which their time scales diverge on the approach to the transition.   The temperature dependence of the critical slowing down is consistent with a continuous transition induced by the freezing out of vortex-like phase fluctuations.
\section{Experimental Details}
Experiments were performed in a broadband Corbino microwave spectrometer. This technique has been used previously to study high-$T_c$ superconductor films, electron glasses, and heavy fermions.\cite{BoothPRL96a,DonghoPRL95a,KitanoPRB09a,MarkleePRL01b,SchefflerNature05}  In this technique, a microwave signal is reflected by a sample that terminates the otherwise open-ended coaxial transmission line and is detected by a network analyzer. The measured complex reflection coefficient, $S_{11}^{m}$,  has the contribution from both the transmission line and sample.  The effects of extraneous reflections, damping, and phase shifts in the transmission line were compensated for by performing three reference measurements on standard samples as described elsewhere.\cite{Scheffler04a,kitano08a}  A blank high-resistivity Si substrate was used as an open standard ($S_{11}=1$).  A 20-nm NiCr film evaporated on Si substrate was used as a load standard (its actual $S_{11}$ can be evaluated from a simultaneous dc measurement). A 20-nm Nb film evaporated on Si substrate was used as a short standard ($S_{11}=-1$ when the Nb film is superconducting).  The sample impedance, $Z_{S}$, can be calculated from the calibrated $S_{11}^{a}$ via the standard expression $Z_{S}=\frac{1+S_{11}^{a}}{1-S_{11}^{a}}Z_{0}$, where $Z_{0}=50$ ohm is the characteristic cable impedance. The substrate contribution is taken into account as described in Ref. \onlinecite{booth96thesis}.  For a thin film, where the sample thickness is much smaller than the skin depth, the complex conductivity is related to sample impedance as $\sigma=\frac{\ln(r_{2}/r_{1})}{2 \pi d Z_{S}}$, where $r_{2}$ and $r_{1}$ are the outer and inner radii of a donut-shaped sample and $d$ is the thickness.  A bias tee allows us to measure the two-contact dc resistance simultaneously with a lock-in amplifier.  This measurement scheme was successfully incorporated into a $^3$He cryostat.  A particular experimental challenge was the heat sinking of the coaxial cable inner conductor.   This was accomplished through  the inclusion of two hermetically sealed glass-bead adapters in the transmission line at the 4.2 K and $^3$He stages separated by a short 10-cm-long superconducting NbTi coaxial cable.  This system can probe the broadband microwave conductivity over the 0.05- to 15-GHz range at temperatures as low as 300 mK.

For these measurements, high-purity (99.999 \%) In$_2$O$_3$ was e-gun evaporated under high vacuum onto clean high-resistivity silicon substrates to a thickness of approximately 30 nm.  Our synthesis methods derive from the work of Ref. \cite{Ovadyahu}, where it was shown that amorphous InO$_x$ can be reproducibly made by a combination of e-beam evaporation of In$_2$O$_3$ with optional annealing.  Essentially similar films have been used in a large number of recent studies of the 2D superconductor-insulator quantum phase transition .\cite{MurthyPRL04a,MurthyPRL05a,Murthy06a,Steiner05a,CranePRB07a,Gantmakher01a}  We believe that the films are morphologically homogeneous with no crystalline inclusions or large-scale morphological disorder because of the following:  TEM-diffraction patterns are diffuse rings with no diffraction spots, AFM images are completely featureless down to a scale of a few nanometers (the resolution of the AFM), and $R$ versus $T$ curves when investigating the 2D superconductor-insulator transition \cite{MurthyPRL04a,CranePRB07a} are smooth with no reentrant behavior that is the hallmark of gross inhomogeneity.  The in-plane penetration depth --- the so called Pearl length ($2 \lambda _{3D} ^2/ d$) \cite{Pearl64a} --- can be calculated from the data below to be approximately 6 mm near $T_c$, which is well in excess of any sample dimension.   Vortices are therefore expected to have logarithmic interactions similar to the case of $^4$He films.

\section{Results}
\begin{figure}[t]
\begin{center}
\includegraphics[width=\columnwidth,angle=0]{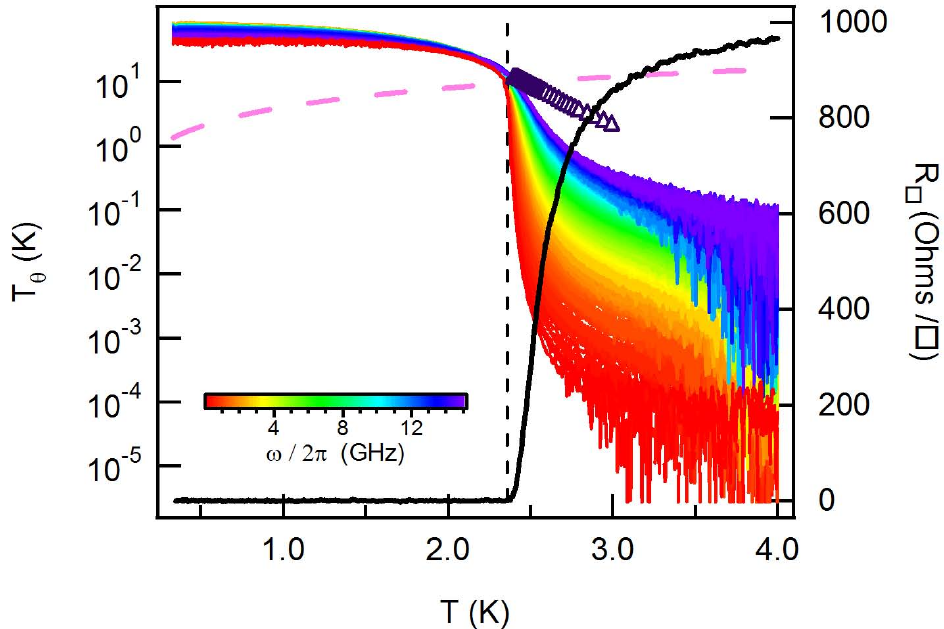}
\caption{(Color online) Temperature dependence of phase stiffness at $\omega/2\pi = 0.21-15$ GHz plotted against the vertical axis on the left.  A color scale representing different frequencies is displayed as well.  The black curve shows resistance per square of the same sample plotted with the vertical axis on the right. The dashed pink line is the KTB prediction, $4 T_{\text{KTB}} = T_{\theta}$, for the universal jump in stiffness. The dark purple $\vartriangle$ markers are $T_{\theta}^{0}$s obtained via the scaling analysis described in the text. $T_c$ is marked by the black dashed line. }
\label{stiffness}
\end{center}
\end{figure}

In this paper, we concentrate on a particular InO$_x$ film with a $T_c = 2.36$ K, but the data is broadly representative of samples with this normal-state resistance.  In what follows, $T_c$ is defined as the temperature at which the simultaneously measured dc resistivity becomes indistinguishable from zero (shown in Fig. \ref{stiffness}).  A small $\pm$ 5 mK uncertainty in this determination does not affect our conclusions.  In Figs. \ref{conductivity} (a) and \ref{conductivity} (b) we plot the real ($G_1 $) and imaginary ($G_2 $) conductance as a function of frequency at different temperatures.  Well above the transition, $G_1$ is flat and featureless and $G_2$ is small, as one expects for a highly disordered metal at low frequencies.  When the sample is cooled toward $T_c$, the real conductance initially becomes enhanced and its spectral weight shifts to lower frequencies.   At lower temperatures, the imaginary conductance grows dramatically and its frequency dependence becomes close to $1 / \omega$.  This is the low-temperature behavior expected for a superconductor.

As seen clearly in plots of the same data as a function of temperature [Fig. \ref{conductivity} (c) and (d)], the region immediately above $T_c$ is dominated by superconducting fluctuations.  As shown by comparison to the dashed line, the real and imaginary conductances begin to show an enhancement in the temperature region above $T_c$.  Our measurements are explicitly sensitive to temporal correlations.   This is seen for instance in the fact that the near-$T_c$ ``dissipation peak" in Fig. \ref{conductivity} (c) is exhibited at lower frequencies for lower temperatures;  the maximum in dissipation is expected when the characteristic fluctuation rate $\Omega/2\pi$ is of the order of the probing frequency $\omega/2\pi$.   The movement of the peak in temperature is a signature of critical slowing down in the raw data.

A particularly important quantity for quantifying fluctuations is the phase stiffness, which is the energy scale to twist the phase of the OP.  The phase stiffness, $T_{\theta}$, is proportional to the superfluid density and can be defined (in units of degrees Kelvin) through the imaginary conductance $G_2$ as

\begin{equation}
k_{B}T_{\theta}(\omega) = \frac{G_{2}}{G_{Q}}\hbar \omega = \frac{N(\omega)e^{2} \hbar d }{m G_{Q}},
\label{stiffdef}
\end{equation}

\noindent where $G_{Q}=\frac{4 e^{2}}{h}$ is the quantum of conductance for Cooper pairs and $N(\omega)$ is a frequency-dependent effective density.  In Fig. \ref{stiffness}, we plot the stiffness versus temperature measured at frequencies between 0.21 to 15 GHz.  $T_{\theta}(\omega)$ defined through Eq. (\ref{stiffdef}) measures the stiffness on a length scale set by the probing frequency, which is typically proportional to the vortex diffusion length during a single radiation cycle, $\sqrt{\frac{\lambda D}{\omega/2\pi}}$.  At temperatures well below $T_c$, there is essentially no frequency dependence to the phase stiffness, consistent with the scenario that the phase stiffness is rigid on all lengths.  At temperatures slightly above $T_c$, the phase stiffness is largest at high frequencies.   In the fluctuation regime, the system retains a phase stiffness on short length scales.  Plotted alongside the stiffness data is the co-measured resistance per square, $R_{\square}$.   Within experimental uncertainty, the phase acquires a frequency dependence at the temperature where the resistance appears to go to zero.  In keeping with our discussion in the introduction, this is reasonable as a superconductor can only exhibit zero resistance when its phase is ordered on all lengths.

KTB theory predicts that at the transition temperature, $T_{\text{KTB}}$, the stiffness in the zero-frequency limit will have a discontinuous jump to zero with a magnitude $T_{\theta} = 4 T_{\text{KTB}} $.  However, because finite frequencies set a length scale, the ac stiffness should go to zero continuously.  We generally expect  that a signature of the discontinuity will manifest in a strong frequency dependence in the stiffness that onsets at $T_{\text{KTB}}$.  In Fig. \ref{stiffness}, the dashed diagonal line gives the prediction \cite{Kosterlitz73a,Halperin79a} for the universal relationship between  $T_{\text{KTB}}$ and the stiffness.   It crosses the stiffness curves very close to where they start to spread.   This, along with the fact that the resistivity goes to zero at this temperature, leads us to assign the transition to a vortex unbinding transition of KTB-like character.  Note that a careful inspection of Fig. \ref{stiffness} on linear scale reveals that the stiffness is in fact approximately 30\% greater at $T_c$ than the universal prediction.  We cannot be sure at this time whether this is a systematic deviation (due perhaps to the dissipative motion of bound vortex pairs or evidence of a nonuniversal jump \cite{Minnhagen87a}) or a small calibration error.

\begin{figure}[t]
\begin{center}
\includegraphics[width=\columnwidth,angle=0]{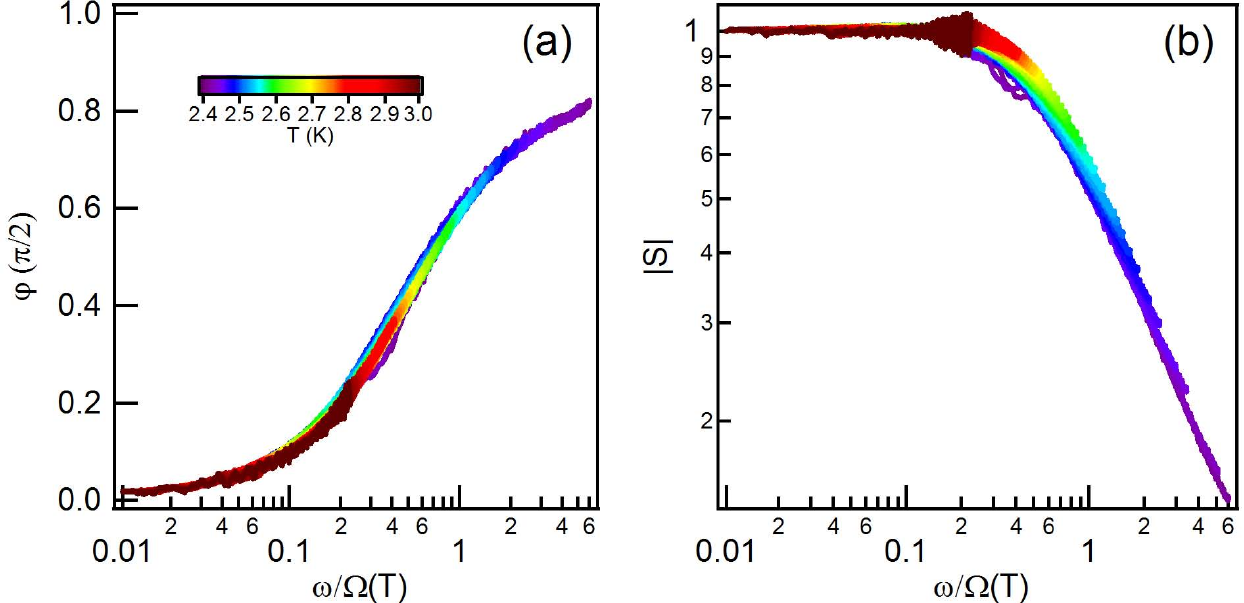}
\caption{(Color online) (a) Phase of $S(\omega/\Omega)$ normalized by $\pi /2$ as a function of reduced frequency $\omega/\Omega$.  (b) Magnitude of $S(\omega/\Omega)$  as a function of reduced frequency. A color scale representing different temperatures for both plots is displayed in (a). Each plot is comprised of data measured at temperatures from 2.398 to 3 K and frequencies from 0.46 to 11 GHz.}
\label{collapse}
\end{center}
\end{figure}

Above $T_{\text{KTB}}$, the conductance due to fluctuating superconductivity is predicted \cite{Fisherfisherhuse,Halperin79a,Corson99a,KitanoPRB09a} to scale with the form
\begin{equation}\label{Scaling}
\frac{G(\omega)} {G_{Q}}=(\frac{k_{B}T_{\theta}^{0}}{\hbar \Omega})S(\omega/\Omega).
\end{equation}
In this scaling function, all temperature dependencies enter through $\Omega$, the characteristic relaxation fluctuation rate, and $T_{\theta}^{0}$,  an overall amplitude factor related to the total spectral weight in the fluctuating part of the conductivity.  Note that in Eq. (\ref{Scaling}) the prefactors $T_{\theta}^{0}$ and $\Omega$ are real quantities, so that the conductance phase angle, $\varphi$, must be equal to the phase angle of $S(\omega/\Omega)$.  By scaling $\omega$ differently for each temperature, we can collapse the phase in a temperature range from 2.398 to 2.993 K into a single universal curve.

In Fig. \ref{collapse} (a), we show this phase angle, $\varphi$, collapsed into a function of reduced frequency $\omega/\Omega$ for each temperature.  The data collapse reasonably well down to 38 mK above $T_c$.  Below 38 mK, the fluctuation frequency begins to enter the low-frequency end of the spectrometer.   In the low scaled-frequency limit, which corresponds to high temperatures and normal-state response, the phase approaches zero as expected.  In the high scaled-frequency limit, which corresponds to low temperatures and the superconducting state, the phase approaches $\pi/2$ also as expected.  This analysis allows us to extract $\Omega(T)$.  Here, we have isolated the fluctuation contribution to the conductance by subtracting off the dc value from well above $T_c$ (at 5.6 K). Having determined $\Omega(T)$, we adjust $T_{\theta}^{0}$ and normalize the magnitude of conductance by $\frac{k_{B}T_{\theta}^{0}}{\hbar\Omega}$ to get the magnitude ($|S|$) of $S(\omega/\Omega)$ so that they fall onto one curve as demonstrated in Fig. \ref{collapse} (b). In 2D, $T_{\theta}^{0}$ is equivalent to the high-frequency limit of the stiffness. We plot it alongside the finite-frequency stiffness in Fig. \ref{stiffness}.

\begin{figure}[t]
\begin{center}
\includegraphics[width=\columnwidth,angle=0]{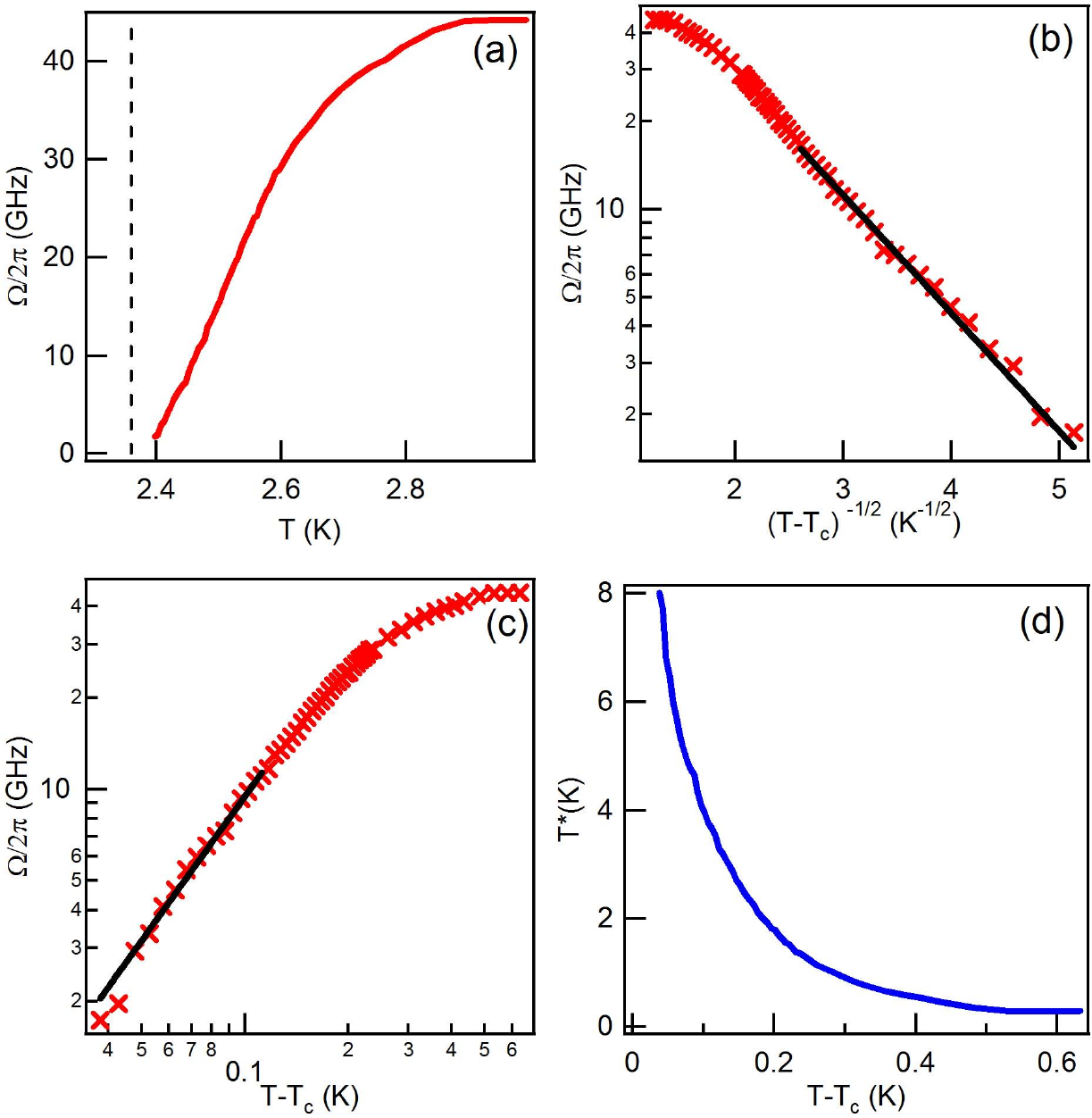}
\caption{(Color online) (a) Fluctuation frequency as a function of temperature from 2.398 to 2.993 K. In (b) and (c), we plot $\Omega(T)/2\pi$ versus $1/\sqrt{T-T_{c}}$ and $T-T_{c}$, respectively, along with the fitting (black curves). (d) Excitation energy in  units of degrees Kelvin as a function of $T-T_{c}$.}
\label{Omega}
\end{center}
\end{figure}

The monotonic decrease of $\Omega(T)$ [Fig. \ref{Omega} (a)] as $T_c$ is approached from above, is an indication for the critical slowing down expected near a continuous transition.  In Figs. \ref{Omega} (b) and \ref{Omega} (c), we fit $\Omega(T)$ to the stretched exponential form expected near a KTB transition $\Omega_0$exp$(- \sqrt{  4T'/(T-T_c)  } )$ as well as to a generic power-law form $\Omega_0 (1 - \frac{T}{T_c} )^{z\nu}$.  The fits were performed over different temperature ranges from $T_c$ on up (147 and 112 mK, respectively) such that the same reduced $\chi^2$ is achieved on both fits.

The stretched-exponential fit gives coefficients of $\Omega_0 /2\pi$ = 181 GHz and $T'= 0.23$ K, while the power-law fit gives $\Omega_0 /2\pi$ = 90 GHz and $z\nu = 1.58$, which are all reasonable parameters.  For instance, within the ansatz of Ref.  \onlinecite{Halperin79a} it is predicted that $T' = \gamma (T_{c0} - T_{\text{KTB}})$, where $\gamma$ is a constant of the order of unity.  Ref. \onlinecite{Benfatto09a} predicts more specifically that $\gamma = 4 \alpha^2$, where $\alpha$ is the ratio of the vortex core energy, $\mu$, to the vortex core energy in the 2D XY model, $\mu_{xy}$.  Viewed in this regard, our value of $T'$ is consistent with a reasonably small value of the core energy.

For both functional forms, one expects that the prefactor $\Omega_0$ will be of order of the inverse time needed to diffuse a vortex with core size $\xi_0$.  Using the Bardeen-Stephen \cite{Halperin79a} approximation for $D$, one can derive the expression $\hbar \Omega_0 = 2 \pi  \lambda \hbar D/\xi_0^2 = 2\pi \lambda \frac{G_Q}{G_N} k_B T_c$, where $\lambda$ is a constant of the order of unity and $G_N$ is the normal-state dc conductance.  For the present sample, this gives $ \Omega_0 / 2\pi \approx\lambda *$48 GHz, which is consistent with both fits.  Due to its larger fitting range and its consistency with the universal jump, we favor the stretched exponential form, but in practice, it is difficult to definitively exclude power-law dependencies.

However, we can note a number of additional aspects consistent with a vortex plasma regime.  Over a more extended temperature range above $T_c$, one expects that $\Omega(T)$ will obey the relation $\Omega_0$exp$( - \frac{T^*}{T})$.   Here, $T^*$ is half the energy needed to thermally excite a free vortex-antivortex pair.  In Fig. \ref{Omega} (d), we plot $T^* = T$ln$(\Omega_0 / \Omega)$.   As expected this quantity appears to diverge as $T \rightarrow T_c$.   It reaches a high-temperature limiting value of about 0.27 K.  One expects \cite{Epstein82a} that $k_B T^* = \mu +  \frac{1}{2} k_B T_{\theta}^{0}$ ln$(\xi/\xi_0)$ as the logarithmic interaction has a cutoff at $\xi$.  At temperatures well above $T_c$ where $\xi$ is of the order of $\xi_0$, the logarithmic term is negligible and the excitation energy should be proportional to the core energy alone.   Within the BCS model, the core energy can be shown  \cite{Epstein82a} to be approximately $k_B T_{\theta}^{0}(T) / 8$.   A comparison with $T_{\theta}^{0}$ from Fig. \ref{stiffness} gives an estimate of $\mu / k_B \approx$ 0.3 K in this temperature range.   The agreement with experiment is essentially exact, but one should not take the exactness too seriously as there are a number of neglected factors of the order of unity.

\section{Conclusion}
We have presented a comprehensive study of the complex microwave conductance of  amorphous superconducting InO$_x$ thin films. Our data explicitly demonstrate critical slowing down close to the phase transition and, in general, the applicability of a vortex-plasma model above $T_c$.  This technique opens up the possibility of studying dynamic scaling of phase transitions at low temperatures and frequencies of a number of material systems.

\section{Acknowledgements}
We thank L. Benfatto, L.S. Bilbro, L. Engel, H. Kitano, M. Scheffler, R. Valdes Aguilar, and L. Zhu for helpful discussions.  The research at JHU was supported by NSF DMR-0847652.   The research at UB was supported by NSF DMR-0847324.

\bibliography{ScIns}
\end{document}